\documentclass[%
 reprint,
showpacs,preprintnumbers,
 amsmath,amssymb,
 aps,
prb,
]{revtex4-2}
\usepackage{xcolor}

\usepackage{graphicx}
\usepackage{dcolumn}
\usepackage{bm}

\newcommand{\milad}[1]{\textcolor{blue}{#1}}
\newcommand{\arman}[1]{\textcolor{magenta}{#1}}
\usepackage[pagebackref=false,colorlinks,linkcolor=blue,filecolor=green,urlcolor=blue ,citecolor=blue]{hyperref}

\usepackage{changes}
\usepackage{cancel}
\usepackage{comment}
\begin{document}

\title{Even-harmonic generation through nonequilibrium steady-state symmetry breaking}

\author{Milad Jangjan}
\affiliation{Institute of Physics, University of Rostock, Albert-Einstein-Stra{\ss}e 23-24, 
D-18059 Rostock, Germany}

\author{Arman Kashef}
\affiliation{Institute of Physics, University of Rostock, Albert-Einstein-Stra{\ss}e 23-24, 
D-18059 Rostock, Germany}

\author{Siamak Pooyan}
\affiliation{Institute of Physics, University of Rostock, Albert-Einstein-Stra{\ss}e 23-24, 
D-18059 Rostock, Germany}

\author{Stefan Scheel}
\affiliation{Institute of Physics, University of Rostock, Albert-Einstein-Stra{\ss}e 23-24, 
D-18059 Rostock, Germany}

\author{Dieter Bauer}
\affiliation{Institute of Physics, University of Rostock, Albert-Einstein-Stra{\ss}e 23-24, 
D-18059 Rostock, Germany}

\begin{abstract}
High-harmonic generation (HHG) in inversion-symmetric systems is typically restricted to odd harmonics by symmetry. Here, we show that this selection rule can be broken without modifying the underlying Hamiltonian. We investigate a boundary-driven Su-Schrieffer-Heeger (SSH) chain coupled to source and sink reservoirs and demonstrate that dissipative dynamics generates a nonequilibrium steady state carrying a finite DC current. While the SSH Hamiltonian retains inversion symmetry, the current-carrying steady-state density matrix does not, leading to the emergence of even harmonics in the emitted spectrum. Using a correlation-matrix approach based on the Lindblad master equation, we obtain the steady state and calculate the resulting HHG response. We find that the intensity of the even harmonics is directly controlled by the transport current, establishing a link between nonequilibrium charge transport and HHG selection rules. Our results uncover a mechanism for even-harmonic generation that relies solely on nonequilibrium steady-state symmetry breaking and provide a route to probing transport currents through ultrafast nonlinear spectroscopy in centrosymmetric quantum systems.
\end{abstract}
\maketitle


\section {Introduction} \label{s1}

Symmetry plays a crucial role in determining the response of quantum systems to external perturbations, and high-harmonic generation (HHG) is no exception. In HHG, strong laser fields drive electrons far from equilibrium, resulting in the emission of radiation at integer multiples of the driving frequency. This process provides a powerful probe of the electronic structure and carrier dynamics in solids~\cite{vampa2015hhg,tancogne2018band}. While HHG studies in the 1990s primarily focused on atomic gases~\cite{McPherson1987,Ferray1988,Krause1992,Corkum1993,Schafer1993,Macklin1993}, the past decade has witnessed rapid progress in solid-state HHG across a broad range of materials and model systems~\cite{Ghimire2011a,Vampa2015,liu2017atomically,Murakami2018,jurss2019ssh_hhg,zhang2024ssh_edge,BauerHansen2018,pooyan2025harmonic,bera2023topological,ghimire2019hhg,cavaletto2025attoscience,PhysRevAMom_Res}.

The harmonic spectrum is strongly constrained by symmetry. In inversion-symmetric systems driven by linearly polarized light, the emitted radiation exhibits half-cycle symmetry, suppressing even harmonics and favoring odd-order harmonics. Consequently, the observation of even harmonics is generally regarded as a signature of inversion-symmetry breaking. However, the relevant symmetry constraints are determined not only by the Hamiltonian but also by the quantum state. A nonequilibrium current-carrying state can break inversion symmetry of the density matrix while the underlying Hamiltonian remains perfectly inversion symmetric. In such a situation, the conventional harmonic selection rules are modified although the lattice itself remains inversion symmetric.

Several mechanisms for generating even harmonics have been identified. In the Su-Schrieffer-Heeger (SSH) model~\cite{SSHModel}, explicit inversion-symmetry breaking leads to the appearance of even harmonics~\cite{Ma2022PRB106125117}. Experimentally, static or THz bias fields applied to centrosymmetric crystals such as silicon induce even harmonics by breaking spatial symmetry~\cite{Ding2022iScience,ThinFilm2023NatComm}. Even-order emission has also been observed in heterostructures such as graphene/boron nitride~\cite{G_hBN2020RSC}, as well as during molecular dissociation where the emission bursts themselves become asymmetric~\cite{SciRepEvenHarmonics}. Another widely used approach employs asymmetric driving fields, such as two-color combinations of the fundamental frequency and its second harmonic~\cite{TwoColor2019ELSPEC}. More recently, it has been suggested that even harmonics may emerge without any structural inversion-symmetry breaking~\cite{kanega2024high,tuovinen2024nanoletters}. In these cases, a finite current induces an asymmetric electronic distribution while the Hamiltonian remains inversion symmetric, thereby modifying the harmonic selection rules.

A persistent current, however, cannot be sustained in a finite isolated system. A realistic description of a current-carrying steady state therefore requires coupling to external reservoirs, naturally placing the problem within the framework of open quantum systems. Lindblad-based descriptions and related dissipative approaches have become standard tools for studying relaxation, decoherence, topology, and nonequilibrium steady states in quantum systems~\cite{nava2023lindblad,yang2023dissipative,yang2022liouvillian,song2019non, PopkovSchutz2017}. In the context of HHG, open-system approaches based on Lindblad dynamics and phenomenological dephasing have been employed to account for relaxation processes during laser-driven dynamics~\cite{floss2024dephasing_hhg}. Furthermore, biased molecular junctions have been shown to exhibit HHG in small current-carrying systems~\cite{tuovinen2024nanoletters}.

Among open quantum systems, the SSH chain provides a particularly attractive platform because it combines nonequilibrium transport with nontrivial topology. Open SSH chains coupled to Lindblad reservoirs or thermal baths have been extensively investigated concerning transport, dissipation, topology, and relaxation dynamics~\cite{klett2018topological,dangel2018topological,guimaraes2016non,yang2023dissipative,nava2023lindblad,yang2022liouvillian,song2019non}. More generally, boundary-driven quantum chains have been shown to support nonequilibrium steady states in particle, spin, and thermal transport settings~\cite{PopkovSchutz2017,he2023particle,Pizorn2013,BoseHubbardBergholtz}. Exact theoretical approaches based on third quantization~\cite{prosen2008third} and correlation-matrix methods~\cite{nava2023lindblad} provide powerful tools for analyzing quadratic fermionic systems, while current-carrying steady states in interacting open quantum chains have also been extensively studied~\cite{NavaRossiGiuliano2021,TarantelliVicari2021}. These works have established a detailed understanding of dissipative transport and nonequilibrium steady states. However, the consequences of such steady currents for high-harmonic generation remain largely unexplored.

To date, most HHG studies consider systems initially prepared in equilibrium. Only a few recent works have begun to investigate HHG in the presence of nonequilibrium steady-state currents. In particular, Ref.~\cite{kanega2024high} demonstrated that a DC current in graphene can enable the generation of even harmonics in addition to the conventional odd harmonics. This raises the fundamental question of how nonequilibrium transport currents modify ultrafast nonlinear optical responses in centrosymmetric systems.

In this work, we address this question by studying HHG in a one-dimensional centrosymmetric SSH chain coupled to source and sink reservoirs. The dissipative dynamics drive the system into a nonequilibrium steady state carrying a finite DC current. Importantly, the SSH Hamiltonian remains inversion symmetric throughout the dynamics. The symmetry breaking responsible for even-harmonic generation arises solely from the nonequilibrium steady-state density matrix induced by the boundary driving. Using an exact correlation-matrix approach, we demonstrate that the resulting transport current enables the appearance of even harmonics without breaking the inversion symmetry of the underlying Hamiltonian.

\begin{figure}[t]
  \centering
\includegraphics[width=0.5\textwidth]{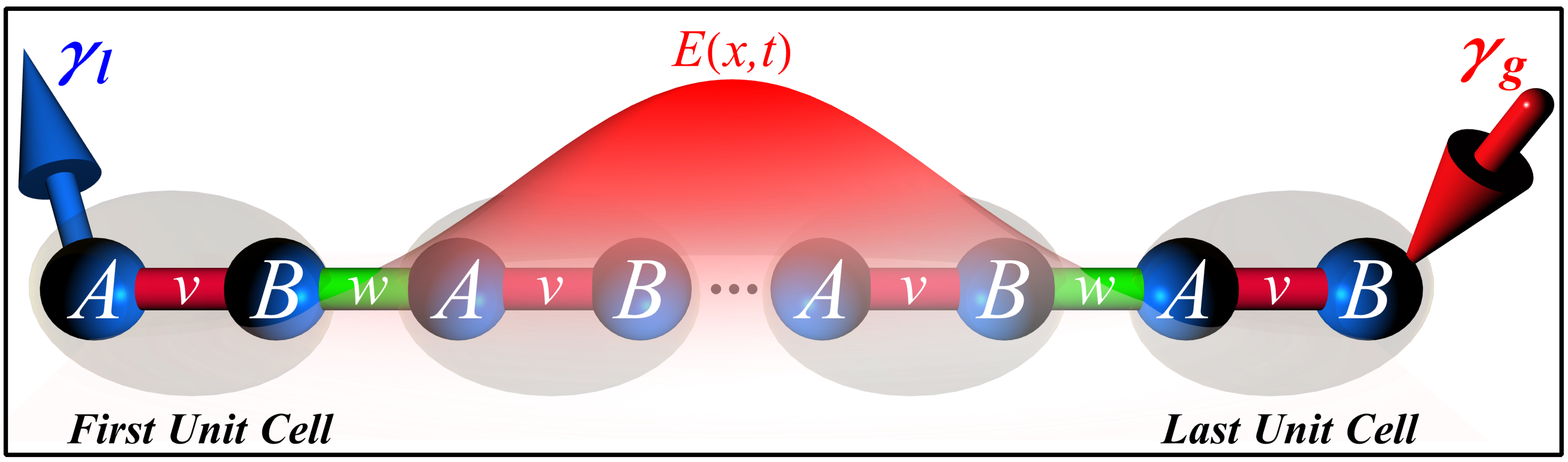}  
  \caption{A 1D SSH chain coupled to a source reservoir (gain rate $\gamma_g$) at the last site and a sink reservoir (loss rate $\gamma_l$) at the first site, driven by a spatiotemporal laser electric field $E(x,t)$.}
  \label{fig:System}
\end{figure}

\section{Model and Theoretical Framework}
In this section, we introduce the one-dimensional SSH model and the open-systems framework used throughout this work. We first describe the SSH Hamiltonian and then formulate the dissipative dynamics within the Lindblad formalism, which drives the system into a nonequilibrium steady state carrying a finite current. Subsequently, a spatially dependent laser field is applied to investigate the resulting high-harmonic response.

\subsection{SSH Hamiltonian}
The one-dimensional SSH model describes electrons on a dimerized lattice with alternating hopping amplitudes between neighboring sites. The Hamiltonian is given by
\begin{eqnarray}
h = -v \sum_{i=1}^{N} c_{2i-1}^{\dagger} c_{2i}
      -w \sum_{i=1}^{N-1} c_{2i+1}^{\dagger} c_{2i}
      + \text{h.c.}
\end{eqnarray}
where $c_m^{(\dagger)}$ is the annihilation (creation) operator of a fermion at site $m$, with odd indices $m=2i-1$ corresponding to sublattice $A$ and even 
indices $m=2i$ corresponding to sublattice $B$ of the $i$th unit cell 
($i=1,\ldots, N$)
as shown in Fig.~\ref{fig:System}. The parameters $v=-\exp [ - (a - 2 \delta)]$ and $w=-\exp [ - (a + 2 \delta)]$ correspond to the intra-cell and inter-cell hopping amplitudes, respectively, and depend exponentially on the dimerization parameter $\delta$. 

\subsection{Dissipative Evolution of SSH chain}

Before studying the laser-driven dynamics, we first determine the nonequilibrium steady state established by the reservoirs. The SSH Hamiltonian introduced above describes the isolated chain. In the absence of external driving, the Hamiltonian $h$ is time-independent. When a time-dependent laser field is applied, the Hamiltonian becomes itself time-dependent, $h(t)$.
In the following, we use $h(t)$ as the general notation for the Hamiltonian, with the understanding that, in the absence of driving, $h(t)\equiv h$. For quadratic Hamiltonians and dissipators that are linear in the fermionic operators, the dynamics of the many-body density matrix $\rho$ can be reduced to a closed set of equations for single-particle observables.

Several approaches have been developed to solve Lindblad dynamics in quadratic fermionic systems, including formulations based on Majorana fermions~\cite{prosen2008third,Lindthird2,lieu2020tenfold,mcdonald2023third,hegde2023edgethird} and complex fermionic operators~\cite{song2019non,NonHermitian2}. In the present work, we adopt the single-particle correlation-matrix representation~\cite{nava2023lindblad,cinnirella2024fate}, which provides an efficient description of the dissipative dynamics and allows direct access to the observables relevant for transport and HHG.

The dissipative dynamics of the system are governed by the Lindblad master equation
\begin{equation}
\begin{aligned}
\dot{\rho}&=\mathcal{L}(t)[\rho]=\mathcal{L}_H(t)[\rho]+\mathcal{L}_D[\rho] \\ \\
&=-i[h(t),\rho(t)]+\sum_k\left(L_k\rho L_k^\dagger-\frac12\{L_k^\dagger L_k,\rho\}\right),
\end{aligned}
\label{eq:lindblad-general}
\end{equation}
where $\mathcal{L}(t)$ is the Liouvillian, $\mathcal{L}_H(t)\rho$ is the coherent unitary evolution generated by the Hamiltonian $h(t)$, and $\mathcal{L}_D$ denotes the dissipative contribution associated with the Lindblad jump operators $L_k$ that accounts for particle exchange with the reservoirs.

For quadratic Hamiltonians and jump operators that are linear in the fermionic operators, the many-body dynamics is closed at the level of the single-particle correlation matrix,
\begin{equation}
\mathcal{C}_{mn}(t)
=
\langle c_m^\dagger c_n\rangle
=
\mathrm{Tr}\!\left(c_m^\dagger c_n\,\rho(t)\right),
\end{equation}
which contains all one-body correlations of the system.

The generic index $k$ labels all Lindblad jump operators. In the following, we separate them into particle-gain and particle-loss operators acting on sites $q$ and $p$, respectively,
\begin{equation}
\label{eq:lindblad-operators}
L_q^{\rm gain}
=
\sqrt{\gamma_g^q}\,c_q^\dagger,
\qquad
L_p^{\rm loss}
=
\sqrt{\gamma_l^p}\,c_p,
\end{equation}
where the indices $q$ and $p$ denote the sites coupled to gain and loss reservoirs, respectively, and $\gamma_g^q$ ($\gamma_l^p$) is the corresponding gain (loss) rate.

Taking the time derivative of $\mathcal{C}_{mn}$, substituting the Lindblad master equation~\eqref{eq:lindblad-general} together with the jump operators into Eq.~\eqref{eq:lindblad-operators}, and using the fermionic anti-commutation relations, we obtain the closed equation of motion for the correlation matrix (see Appendix~\ref{App_DensityMatrix}),
\begin{eqnarray}
\dot{\mathcal{C}}
&=&
i[h(t),\mathcal{C}(t)]-\frac12\sum_p\left(\Gamma_p^-\mathcal{C}(t)+\mathcal{C}(t)\Gamma_p^-\right)\nonumber\\
&+&\sum_q\left[\Gamma_q^+-\frac12
\left(\Gamma_q^+\mathcal{C}(t)+\mathcal{C}(t)\Gamma_q^+
\right)\right],
\label{MainEq}
\end{eqnarray}
where
\[
\Gamma_{p(q)}^{-(+)}=\gamma_{l(g)}^{p(q)}P_{p(q)}=\gamma_{l(g)}^{p(q)}|p(q)\rangle\langle p(q)|
\]
describes local particle loss (gain) at site $p$ ($q$) with rate $\gamma_l^p$ ($\gamma_g^q$). Here, $|j\rangle$ ($j=1,\ldots,2N$) denotes the canonical site basis of the single-particle Hilbert space, satisfying $\langle i|j\rangle=\delta_{ij}$, so that $P_j=|j\rangle\langle j|$ projects onto site $j$.


Throughout this work, we consider particle loss at the first site, $p=1$, and particle gain at the last site, $q=2N$, with equal coupling strengths $\gamma_l=\gamma_g=\gamma$. This boundary driving establishes a finite nonequilibrium steady-state current across the system. Collecting terms in Eq.~\eqref{MainEq}, the equation of motion can be written in the compact form
\begin{equation}
\dot{\mathcal C}=\mathcal M(t)\mathcal C(t)+\mathcal C(t)\mathcal N(t)+\Gamma^+,
\label{eq:Close_form_timeSS}
\end{equation}
where
\begin{align}
\mathcal M&=i h(t)-\frac{\Gamma}{2},\qquad\mathcal N
=\mathcal M^\dagger,\nonumber \\ \nonumber
\Gamma&=\Gamma^++\Gamma^-.
\end{align}

Equation~\eqref{eq:Close_form_timeSS} forms the basis of our analysis. In the absence of laser driving, it determines the nonequilibrium steady state generated by the reservoirs, while in the presence of a time-dependent field it governs the subsequent dynamics used to calculate the current and the HHG spectra.

\subsection{Steady-State Solution }\label{sec}
At sufficiently long times, provided the dynamics is stable, the correlation matrix approaches a non-equilibrium steady state. Formally, the steady state is defined by
\begin{equation}
\mathcal{C}_\mathrm{ss} = \lim_{t \to \infty} \mathcal{C}(t), 
\qquad \text{with} \qquad
\frac{d\mathcal{C}}{dt} = 0.
\label{ConditionSS}
\end{equation}

Substituting the steady-state condition into Eq.~\eqref{MainEq} yields
\begin{equation}
\mathcal{M} \mathcal{C}_\mathrm{ss} + \mathcal{C}_\mathrm{ss} \mathcal{N} = - \Gamma^+,
\label{eq:sylvester_form}
\end{equation}
where $\mathcal{M}(t)\equiv \mathcal{M}$ and $\mathcal{N}(t)\equiv \mathcal{N}$ because the SSH Hamiltonian is time independent. In general, Eq.~\eqref{eq:sylvester_form} is a Sylvester equation~\cite{bartels1972solution}. A unique solution exists if and only if the spectra of $\mathcal{M}$ and $-\mathcal{N}$ do not overlap~\cite{bartels1972solution}. Moreover, for this model $\mathcal{M}=\mathcal{N}^{\dagger}$ and Eq.~\eqref{eq:sylvester_form} reduces to a Lyapunov equation.

In principle, the steady-state correlation matrix can also be obtained by integrating Eq.~\eqref{MainEq} and evolving the system to sufficiently long times. This dynamical approach avoids solving the steady-state equation explicitly and instead extracts the steady state from the long-time limit of the evolution.

An analytical solution of Eq.~\eqref{eq:sylvester_form} can be obtained as
\begin{equation}
    \mathcal{C}_{\rm ss} = \int_0^\infty e^{\mathcal{M}t} \Gamma^+ e^{\mathcal{M}^\dagger t} dt.
\end{equation}
This integral converges provided all eigenvalues of $\mathcal{M}$ have negative real parts. Expanding the solution in the eigenbasis of $\mathcal{M}$ yields~\cite{wang2024dynamic} (see Appendix~\ref{AppSS})
\begin{equation}
\mathcal{C}_{\rm ss} = \gamma \sum_{k,\ell} \frac{\langle\tilde u_k|2N\rangle \langle 2N|\tilde u_\ell\rangle^*}{-(\lambda_k+\lambda_\ell^*)}\; |u_k\rangle\langle u_\ell|,
\label{eq:SSAnalitycal}
\end{equation}
%
where $|\tilde u_k\rangle$ ($|u_k\rangle$) and $\lambda_k$ denote the left (right) eigenvectors and eigenvalues of $\mathcal{M}$, respectively.
The approach to the steady state is governed by the spectrum of the matrix $\mathcal{M}$ in Eq.~\eqref{eq:Close_form_timeSS}. Its eigenvalues determine the relaxation rates through their real parts.~\cite{song2019non,van2018symmetry,cai2013algebraic}. The corresponding Liouvillian gap is defined as $\Lambda = \min(2\mathrm{Re}(-\lambda))$.
When this gap is finite, the system relaxes to a unique steady state on a finite timescale, and the long-time dynamics become independent of the initial conditions. However, the gap depends on the system parameters and may close in certain regimes. In particular, in the limit $\gamma \rightarrow 0$, the gap can vanish, leading to a divergence of the relaxation time and rendering the steady-state limit singular. In such cases, the convergence towards the steady state is generally algebraic rather than exponential~\cite{song2019non}.

\subsection{Driven Dynamics under a Spatially and Temporally Dependent Electric Field}
The positions of the atomic sites in the SSH chain are given by
\begin{eqnarray}
x_j = \left( j - \frac{2N + 1}{2} \right)a - (-1)^j \delta, \ \ j = 1,2,\dots,2N,
\end{eqnarray}
where $a$ denotes the lattice constant and $\delta$ is the dimerization parameter responsible for the alternating displacement of neighboring sites.

As shown in Fig.~\ref{fig:System}, we employ a spatially and temporally dependent electric field to investigate HHG in the current-carrying SSH chain. We introduce the spatial modulation in order to avoid illuminating the source and sink sites. The interaction between the electrons and the external laser field can still be described in dipole approximation and  length gauge. The corresponding light-matter interaction Hamiltonian is given by
\begin{eqnarray}
\mathfrak{h}_{\text{laser}}(t) = -\sum_{j} E(x_j,t)d_j \  c_{j}^\dagger c_{j},
\end{eqnarray}
where $E(x,t)$ denotes the external electric field and $d_j$ is the local dipole moment operator. For an electron localized at position $x_j$, the dipole moment is proportional to its position,
\begin{eqnarray}
d_j = e x_j,
\end{eqnarray}
where the electron charge is taken to be $e=-1$.
The total time-dependent Hamiltonian of the driven SSH system is then given by
\begin{eqnarray}
h(t) = h + \mathfrak{h}_{\text{laser}}(t),
\end{eqnarray}
which describes the dynamics of the SSH chain in the presence of the external laser field. This Hamiltonian serves as the starting point for the time-dependent Lindblad evolution used to calculate the laser-induced current and the resulting high-harmonic spectra.

\section{High-Harmonic Generation}
To investigate the nonlinear optical response of the nonequilibrium steady state, the SSH chain is driven by a spatiotemporal laser pulse of the form
\begin{eqnarray}
E(x,t) = S(x) E(t),
\quad 0 < t < n_{\text{cyc}}\frac{2\pi}{\omega},
\end{eqnarray}
where $E(t)$ is the temporal profile of the pulse and $S(x)$ its spatial envelope,
\begin{eqnarray}
   E(t)= E_0 \sin^2\!\left(\frac{\omega t}{2 n_{\text{cyc}}}\right)\cos(\omega t), \nonumber \\
  S(x_j) = e^{-\frac{(x-x_0) ^2}{2\sigma^2}}\, 
          0.5\!\left(1 - \cos\left(\frac{2 \pi  j}{2N - 1}\right)\right).
\end{eqnarray}
Here, $E_0$ denotes the peak field amplitude, $\omega$ the carrier frequency, and $n_{\text{cyc}}$ the number of optical cycles in the pulse. The temporal profile is chosen as a $\sin^2$-enveloped pulse, while the spatial profile consists of a Gaussian envelope of width $\sigma = 2N/(2\sqrt{2\ln 2})$ centered at $x_0$ and a Hann window. The latter yields zero electric field at the boundary sites coupled to the source and sink reservoirs, thereby avoiding direct driving of the dissipative contacts.

The HHG spectrum can be evaluated using equivalent formulations based on the dipole moment, dipole acceleration, or current response of the system~\cite{bandrauk2009quantum,baggesen2011dipole}. In the present work, we calculate the emitted spectrum from the current. The corresponding power spectrum is defined as
\begin{eqnarray}
P_J(\omega) \propto 
\left| \mathrm{FFT}\left[\langle J(t) \rangle\right] \right|^2,
\end{eqnarray}
where 
$\langle \hat{J}(t) \rangle=-2 \sum_{i}\operatorname{Im}\!\left[t_i\,\mathcal{C}_{i,i+1}(t)\right]$,
and $t_i$ denotes the hopping amplitude on the corresponding bond, taking the values $v$ or $w$ depending on whether the bond is intra-cell or inter-cell.

\section{Selection Rules from the Liouvillian Structure}
In closed quantum systems, harmonic selection rules are determined by the symmetries of the Hamiltonian and the driving field~\cite{Alon1998SelectionRules,Ceccherini2001DynamicalSymmetry,Neufeld2019FloquetSelection}. In open quantum systems, however, the relevant symmetry is that of the full Liouvillian superoperator rather than the Hamiltonian alone. This distinction is particularly important for nonequilibrium steady states generated by dissipative dynamics. In the present system, the SSH Hamiltonian remains inversion symmetric, while the coupling to source and sink reservoirs generates a finite dc current. As a consequence, the steady-state density matrix need not inherit the inversion symmetry of the Hamiltonian, leading to modified harmonic selection rules.

\begin{figure}[t!]
\centering
\includegraphics[width=0.5 \textwidth]{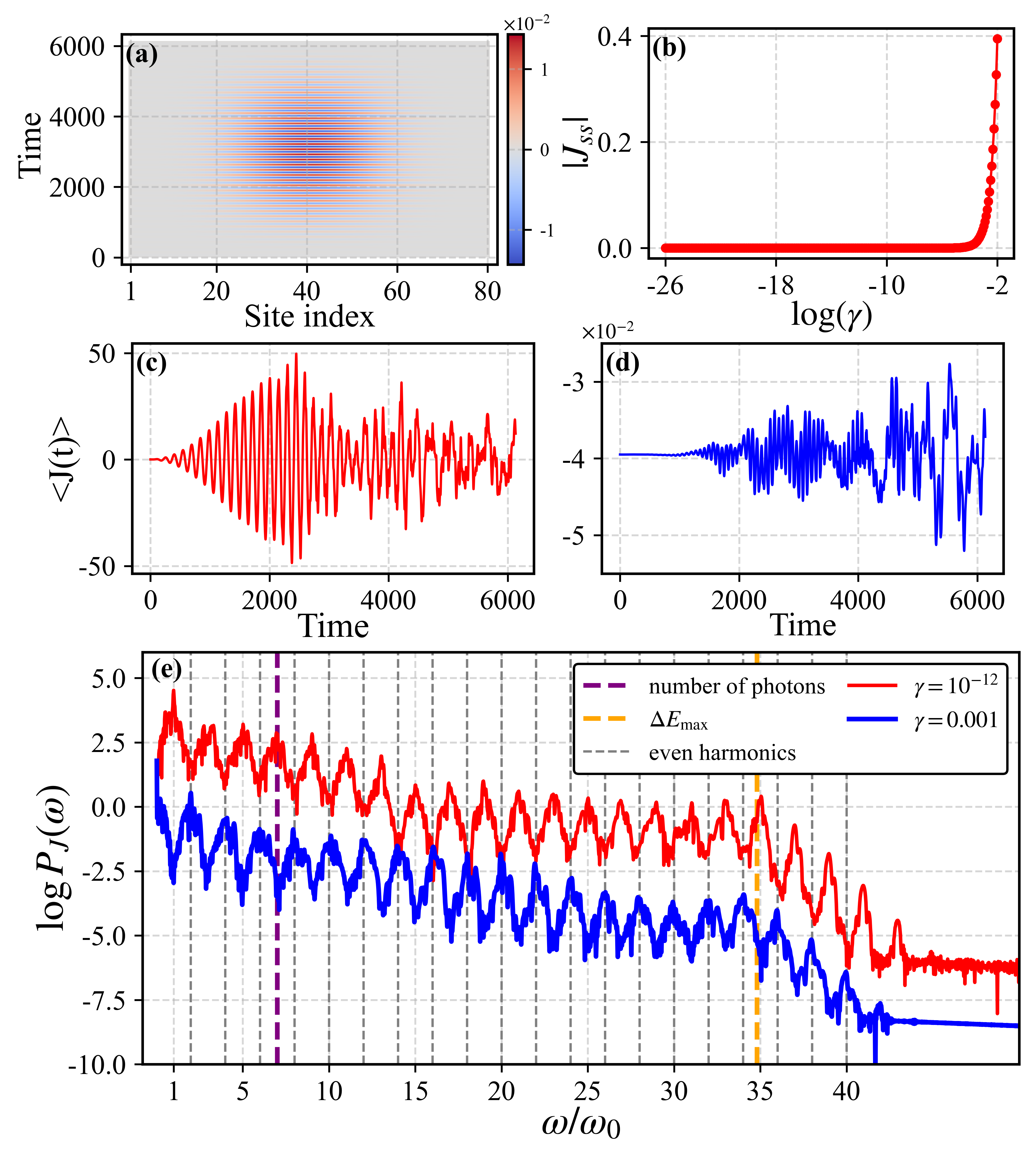}
\caption{ (a) Applied electric field in time and space. (b) Steady-state current as a function of source and sink strength. Total current expectation values as a function of time for (c) $\gamma=10^{-12}$ and (d) $\gamma=0.001$. (e) Corresponding HH spectra; calculations consider 7 photons in the spectral gap. Here,  $\delta = 0.1$, $\omega_0 = \Delta E/7$, $a_0 = 0.5$, and $n_\mathrm{cyc} = 42$, $N=40$. 
}
\label{fig:HHG_Current}
\end{figure}

%
In the absence of dissipation ($\gamma = 0$), the Liouvillian reduces to the purely unitary generator described by the von Neumann equation
\begin{equation}\label{eq:Coh Lio}
\mathcal{L}_H(t)[ \rho] = -i[h(t),\rho].
\end{equation}

The SSH Hamiltonian possesses inversion symmetry, $\mathcal{P} h \mathcal{P}^{-1} = h$, while the driving field satisfies half-cycle symmetry, $h(t+T/2) = -h(t)$. Together, these symmetries imply $\mathcal{P} h(t) \mathcal{P}^{-1} = h(t+T/2)$.For the finite $\sin^2$-enveloped pulse used in our simulations, this symmetry holds only approximately, with corrections of order $1/n_{\rm cyc}$, which are negligible for $n_{\rm cyc}=42$. Since conjugation by $\mathcal{P}$ passes through the commutator, the coherent Liouvillian, i.e., Eq.~\ref{eq:Coh Lio} satisfies
$$
\mathcal{P}\mathcal{L}_H(t)\mathcal{P}^{-1}=\mathcal{L}_H(t+T/2).
$$ Provided the ground state is non-degenerate, the equilibrium density matrix also respects inversion symmetry,
$\mathcal{P} \rho_0 \mathcal{P}^{-1} = \rho_0$.
Since $\mathcal{P}\rho(t)\mathcal{P}^{-1}$ obeys the same equation of motion and initial condition as $\rho(t+T/2)$, uniqueness of the Liouville equation implies
$$
\mathcal{P}\rho(t)\mathcal{P}^{-1}=\rho(t+T/2).
$$
Finally, because $
J(t)=\operatorname{Tr}[\rho(t)J]
=\operatorname{Tr}[(\mathcal{P}\rho(t)\mathcal{P}^{-1})(\mathcal{P}J\mathcal{P}^{-1})]
\nonumber 
=\operatorname{Tr}[\rho(t+T/2)(-J)]
=-J(t+T/2)
$
we have
\begin{equation}\label{eq: J}
J(t+T/2) = -J(t).
\end{equation}
Using Eq.(\ref{eq: J}) and the Fourier series,
\begin{eqnarray}
J(t+T/2) = \sum_n J_n\,e^{-in\omega(t+T/2)}\\ \nonumber
            =
            \sum_n J_n\,(-1)^n\,e^{-in\omega t},
\end{eqnarray}
matching coefficients with the corresponding Fourier series for $-J(t)$ gives $J_n(-1)^n = -J_n$, which requires $(-1)^n = -1$, i.e., $n$ is odd. Therefore
\begin{equation}
J_n = 0 \quad \text{for even } n,
\end{equation}
which enforces the appearance of odd harmonics only. The situation changes once the chain is coupled to source and sink reservoirs. The jump operators introduced in Eq.~(\ref{eq:lindblad-operators}) explicitly break inversion symmetry,
\begin{equation}
\mathcal{P} L_{\text{gain}} \mathcal{P}^{-1} 
\neq L_{\text{loss}}.
\end{equation}
and therefore
\begin{equation}
\mathcal{P} \mathcal{L}_D \mathcal{P}^{-1} 
\neq \mathcal{L}_D.
\end{equation}
As a consequence, the nonequilibrium steady-state density matrix no longer satisfies inversion symmetry,
\begin{equation}
\mathcal{P} \rho_{\mathrm{ss}} \mathcal{P}^{-1} 
\neq \rho_{\mathrm{ss}}.
\end{equation}
The resulting finite steady-state current, $J_{\mathrm{dc}} \neq 0$, shifts the laser-induced current oscillations according to $J(t)=J_{\mathrm{dc}}+J_{\mathrm{ac}}(t)$. Even if $J_{\mathrm{ac}}(t)$ approximately retains half-cycle symmetry, the finite dc offset destroys the exact antisymmetry condition required for suppressing even harmonics. The corresponding selection rule is therefore lifted.

Consequently, the emergence of even harmonics does not require structural inversion-symmetry breaking of the lattice. For an inversion-symmetric Hamiltonian, symmetric driving field, and symmetric initial state, the appearance of even harmonics must originate from the dissipative dynamics. In the present case, the symmetry breaking occurs at the level of the Liouvillian through the boundary gain and loss processes. These processes generate a nonequilibrium steady state whose density matrix no longer obeys inversion symmetry, thereby allowing even-order harmonic emission.

\begin{figure}[t!]
\centering
\includegraphics[width=0.5\textwidth]{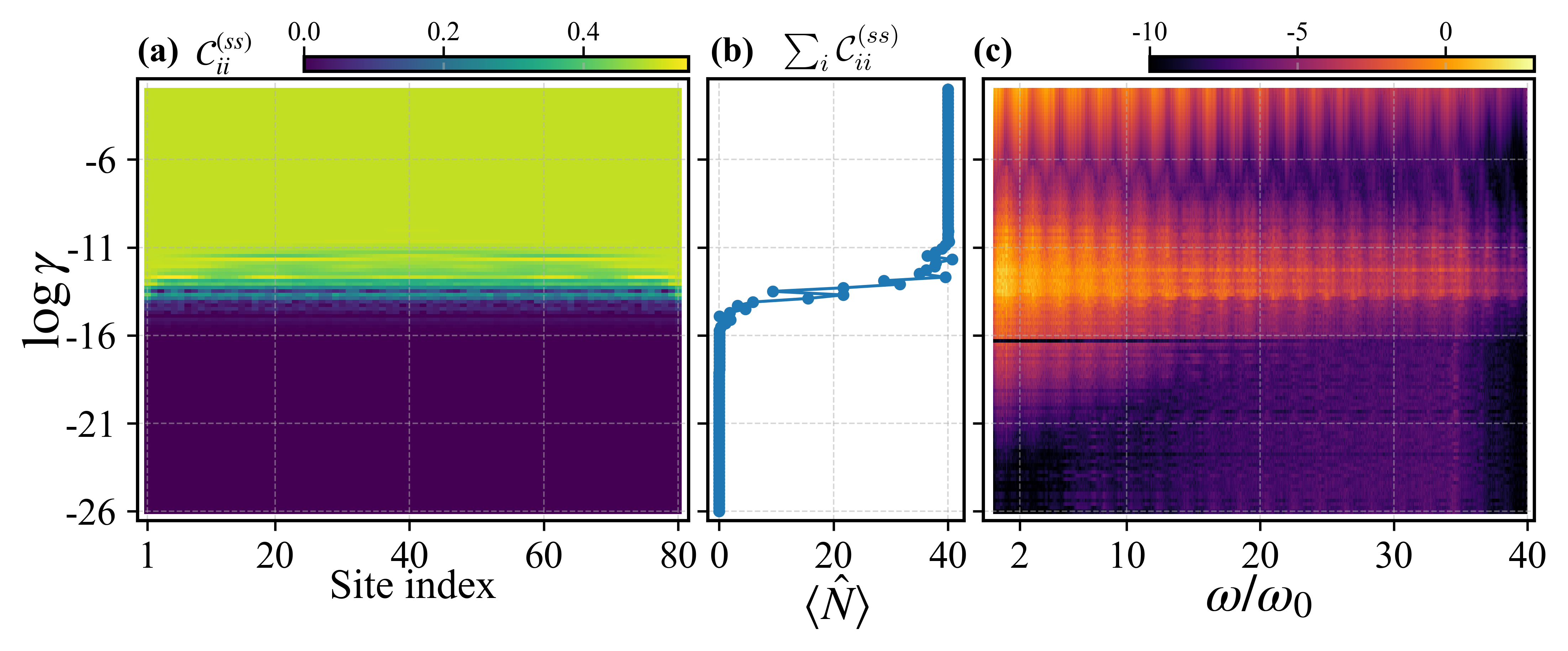}
\caption{(a) Particle number vs.\ site index and $\gamma$. (b) Total number of particles as a function of $\gamma$.  (c) Contour plot of harmonic yields as a function of harmonic order and $\gamma$.  
}
\label{fig:Diagram}
\end{figure}

\section{Results and Discussion} \label{s8}

Figure~\ref{fig:HHG_Current} shows the high-harmonic response of the current-carrying SSH chain. The driving frequency is chosen as $\omega_0=\Delta E/7=0.043$, and the peak electric-field amplitude is $E_0=A_0\omega=0.0172$ with the number of cycles $n_\mathrm{cyc}=42$. $\Delta E=0.297$ is the energy gap of the undriven SSH chain. In addition, the number of unit cells and the dimerization parameter are $N=40$ and $\delta=0.1$, respectively. Finite-size scaling analysis confirms that the emergence of even harmonics in the HHG spectrum is independent of system size in the large-$N$ regime. Panel (a) shows the spatiotemporal profile of the applied laser field $E(x,t)$. The field is localized around the center of the chain. Consequently, the laser primarily interacts with the bulk of the system without directly driving the sites connected to the source and sink reservoirs.

The steady-state current is shown as a function of the dissipation strength $\gamma$ in Fig.~\ref{fig:HHG_Current}(b). As the gain and loss rates increase, the magnitude of the steady-state current increases correspondingly, reflecting the enhanced particle transport induced by the reservoirs.

The time-dependent expectation value of the total current is displayed in Fig.~\ref{fig:HHG_Current}(c,d). For $\gamma=10^{-12}$, the dynamics are effectively governed by the von Neumann equation and therefore retain the half-cycle symmetry of the driven SSH Hamiltonian. In contrast, for finite dissipation, $\gamma=0.001$, the system supports a nonzero steady-state current even before the laser pulse is applied. This current is obtained from Eqs.~\ref{eq:sylvester_form},\ref{eq:SSAnalitycal} and appears as a finite offset in the current signal. Once the laser field is switched on, the induced oscillations occur around this nonequilibrium steady-state current rather than around zero. The suppressed oscillation amplitude at finite $\gamma$ arises because the boundary dissipators continuously damp the laser-induced bond coherences $\mathcal{C}_{i,i+1}(t)$. The coherent drive therefore competes directly with dissipative decay.

The corresponding HHG spectra are shown in Fig.~\ref{fig:HHG_Current}(e). The gray, purple, and orange dashed lines indicate the positions of the even harmonic orders, the band gap, and the energy difference between the lowest and highest single-particle SSH eigenstates, respectively. The red and blue curves correspond to the current traces shown in Fig.~\ref{fig:HHG_Current}(c,d). Harmonics below the band gap $\Delta E$ are associated with intraband dynamics whereas harmonics above contain  interband contributions. Because of $7\omega=\Delta E$,  the 7th harmonic marks the onset of   interband excitation within the SSH energy spectrum.

For $\gamma=10^{-12}$, the spectrum is dominated by odd harmonics, consistent with the conventional selection rules of inversion-symmetric systems. In this limit, the influence of the jump operators becomes negligible and the system behaves effectively as a closed chain. For finite dissipation, $\gamma=0.001$, even-order harmonics emerge and substantially modify the low-order HHG spectrum. Their appearance is not caused by structural inversion-symmetry breaking of the lattice, but instead originates from the nonequilibrium steady-state current generated by the dissipative boundary driving. The steady current also redistributes spectral weight among the harmonic orders, reflecting the interplay between coherent laser-driven dynamics and dissipative transport.

Figure~\ref{fig:Diagram} presents the steady-state particle distribution, the total particle number, and the HHG spectrum as a function of the dissipation strength $\gamma$. The steady-state particle density exhibits a pronounced dependence on the reservoir coupling strength. [see Fig.~\ref{fig:Diagram}(a,b).]

At very small $\gamma$ ($\gamma \lesssim 10^{-16}$), the reservoirs couple too weakly to establish any appreciable steady-state occupation: the Liouvillian gap $\Lambda \sim \gamma$ vanishes, and the net particle flux sustained by the source and sink becomes negligible. The steady state therefore approaches the trivial vacuum,
$$
\mathrm{Tr}[\mathcal{C}_{\mathrm{ss}}]\rightarrow 0,
$$
as discussed further below.
For intermediate coupling,
$
10^{-16}\lesssim \gamma \lesssim 10^{-11},
$
the system crosses over between the vacuum and the fully populated chain: the total particle number
$
\mathrm{Tr}[\mathcal{C}_{\mathrm{ss}}]
$
grows smoothly from $0$ toward $N$ as the reservoirs become progressively more effective at injecting and extracting particles.
At large $\gamma$ ($\gamma \gtrsim 10^{-11}$), dissipative exchange becomes fast compared with the internal relaxation timescale. Coherent hopping redistributes each injected particle across the chain before the next exchange event. In this regime, the steady-state correlation matrix approaches the featureless, maximum-entropy form corresponding to uniform half-filling. Since $\mathcal{C}_{\mathrm{ss}}$ is diagonal with all entries equal to $1/2$,
$$
\mathrm{Tr}[\mathcal{C}_{\mathrm{ss}}]=2N\times\frac{1}{2}=N,
$$
i.e., the chain is on average half-filled.

As shown in Fig.~\ref{fig:Diagram}(c), this regime coincides with the appearance of even harmonics in the HHG spectrum, demonstrating the crucial role of the pre-existing nonequilibrium current before the laser pulse is applied. As $\gamma$ decreases, the boundary driving becomes progressively weaker and the dissipative timescale $t_{\mathrm{diss}}\sim1/\gamma$ increases. The dynamics then become increasingly dominated by coherent Hamiltonian evolution. Consequently, the influence of the jump operators on the symmetry properties of the system is reduced, the nonequilibrium current decreases, and the HHG spectrum gradually recovers the odd-harmonic structure characteristic of inversion-symmetric systems. It is important to emphasize that the vacuum limit refers to the nonequilibrium steady state of the open system. It does not describe the equilibrium state of the isolated SSH Hamiltonian. Although one might expect the system to recover the SSH ground state as $\gamma \to 0$, the open-system steady state instead becomes trivial, since no source remains to maintain a finite particle population once the reservoir coupling vanishes. Moreover, in this regime the convergence towards the steady state becomes algebraic rather than exponential, reflecting the vanishing Liouvillian gap~\cite{song2019non}.

The exact steady-state results presented in Fig.~\ref{fig:HHG_Current} and Fig.~\ref{fig:Diagram} are obtained from the Sylvester equation. While this approach provides direct access to the nonequilibrium steady state, experimentally such a state is reached through a finite-time relaxation process. A more realistic preparation protocol therefore consists of initializing the SSH chain and allowing it to evolve under the Lindblad dynamics for a relaxation time $\tau$ before the laser pulse is applied. Starting from the half-filled SSH state, this procedure enables us to investigate how the approach to the steady state influences the resulting HHG response.

\begin{figure}[t!]
    \centering
    \includegraphics[width=0.5 \textwidth]{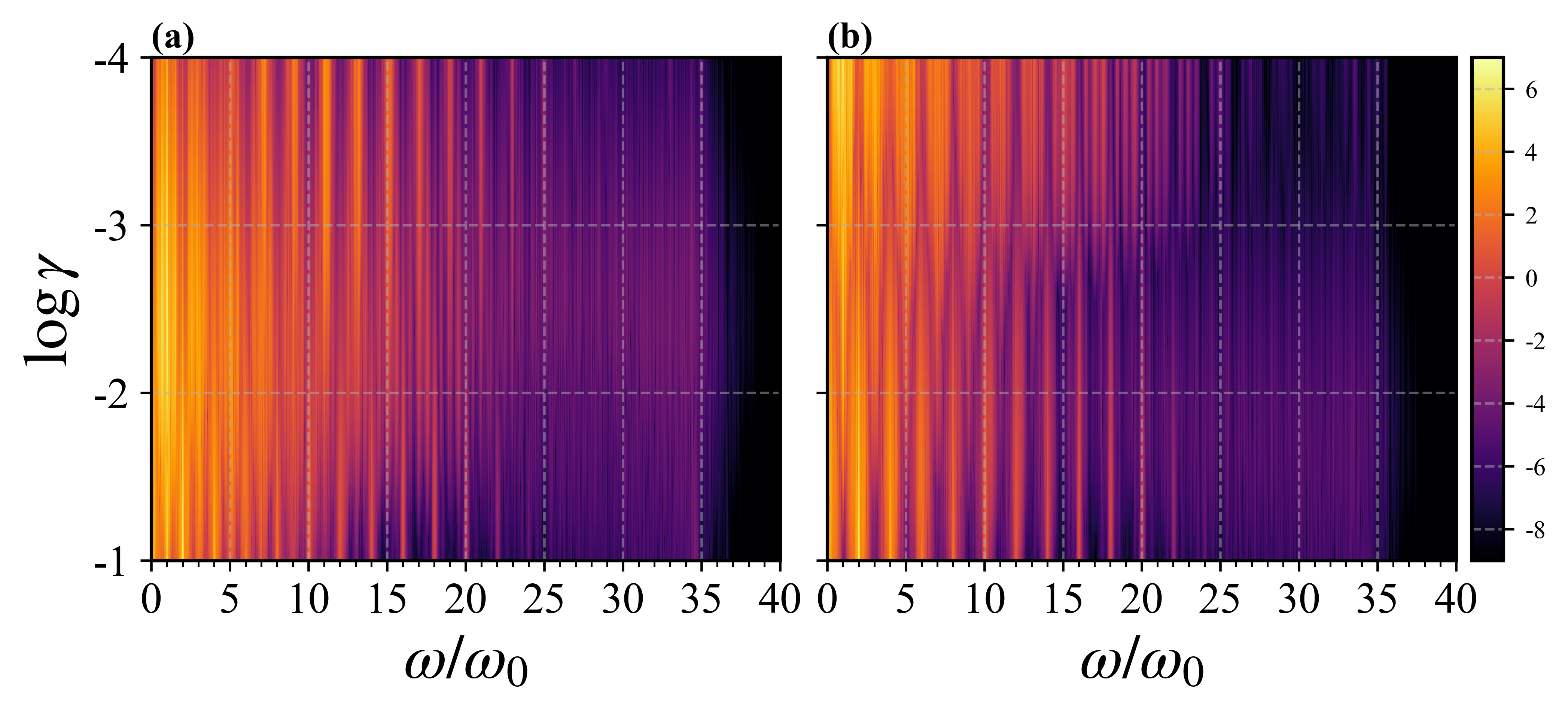}  
    \caption{Harmonic spectra as a function of $\log\gamma$  for (a) $\tau=10^6$ and (b) $\tau=10^7$.
    }
    \label{fig:Diagram2}
\end{figure}

Figure~\ref{fig:Diagram2} presents contour plots of the harmonic yield as a function of harmonic order and dissipation strength $\gamma$ for two different relaxation times, $\tau=10^{6}$ and $\tau=10^{7}$, shown in panels (a) and (b), respectively. The results demonstrate that the emergence of even harmonics depends not only on the reservoir coupling strength but also on the available relaxation time. For the shorter evolution time, $\tau=10^{6}$, the parameter region in which even harmonics are visible is relatively narrow. Increasing the relaxation time to $\tau=10^{7}$ significantly expands this region, indicating that the system requires progressively longer times to develop the current-carrying steady state in the weak-dissipation regime.

For small $\gamma$, the Liouvillian gap becomes small and the relaxation towards the nonequilibrium steady state is correspondingly slow. If the system is not evolved for a sufficiently long time, it retains a partial memory of its inversion-symmetric initial state, $\rho_\mathrm{SSH}$, thereby suppressing the appearance of even harmonics. Increasing the relaxation time allows the system to approach the current-carrying steady state more closely, enhancing the inversion-symmetry breaking of the density matrix and promoting even-harmonic generation.

This dynamical preparation protocol differs conceptually from the Sylvester-equation approach employed above. The Sylvester equation directly yields the exact steady-state correlation matrix, independent of the preparation pathway, and therefore represents an ideal steady-state description. In contrast, the time-dependent Lindblad evolution explicitly captures the transient relaxation process through which the steady state is established. From an experimental perspective, the latter provides a more realistic description, since nonequilibrium steady states are reached through finite-time relaxation, e.g., starting from the isolated system's ground state,  rather than by an instantaneous projection onto the steady-state manifold.

To quantify the convergence towards the steady state, we compute the fidelity between the time-evolved correlation matrix $\mathcal{C}(\tau)$ and the steady-state matrix $\mathcal{C}_{\mathrm{ss}}$ obtained from the Sylvester equation. As the Hamiltonian is quadratic and the Lindblad operators are linear in the fermionic operators, the system remains within the class of Gaussian states~\cite{carollo2018uhlmann}. Consequently, the many-particle density matrix is fully determined by the single-particle correlation matrix $\mathcal{C}$~\cite{carollo2018uhlmann}. As shown in Appendix~\ref{App:Fidelity}, the fidelity can therefore be evaluated directly from the corresponding correlation matrices as
\begin{equation}
\begin{aligned}
F(t)
&\equiv
F\left(\rho(t),\rho_{\mathrm{ss}}\right)=
\det\Big[(\mathbf1-\mathcal C(t))
(\mathbf1-\mathcal C_{\mathrm{ss}})\Big]\\
&\times \det\left[
\mathbf1+
\left(
\sqrt{X(t)}
X_{\mathrm{ss}}
\sqrt{X(t)}
\right)^{1/2}
\right]^2,
\end{aligned}
\label{eq:fidelity}
\end{equation}
where
\[
X(t)
=
\mathcal C(t)
(\mathbf1-\mathcal C(t))^{-1},
\qquad
X_{\mathrm{ss}}
=
\mathcal C_{\mathrm{ss}}
(\mathbf1-\mathcal C_{\mathrm{ss}})^{-1}
\]
with $F(t)\to 1$ signalling convergence to the nonequilibrium steady state.

\begin{figure}[t!]
    \centering
    \includegraphics[width=0.4 \textwidth]{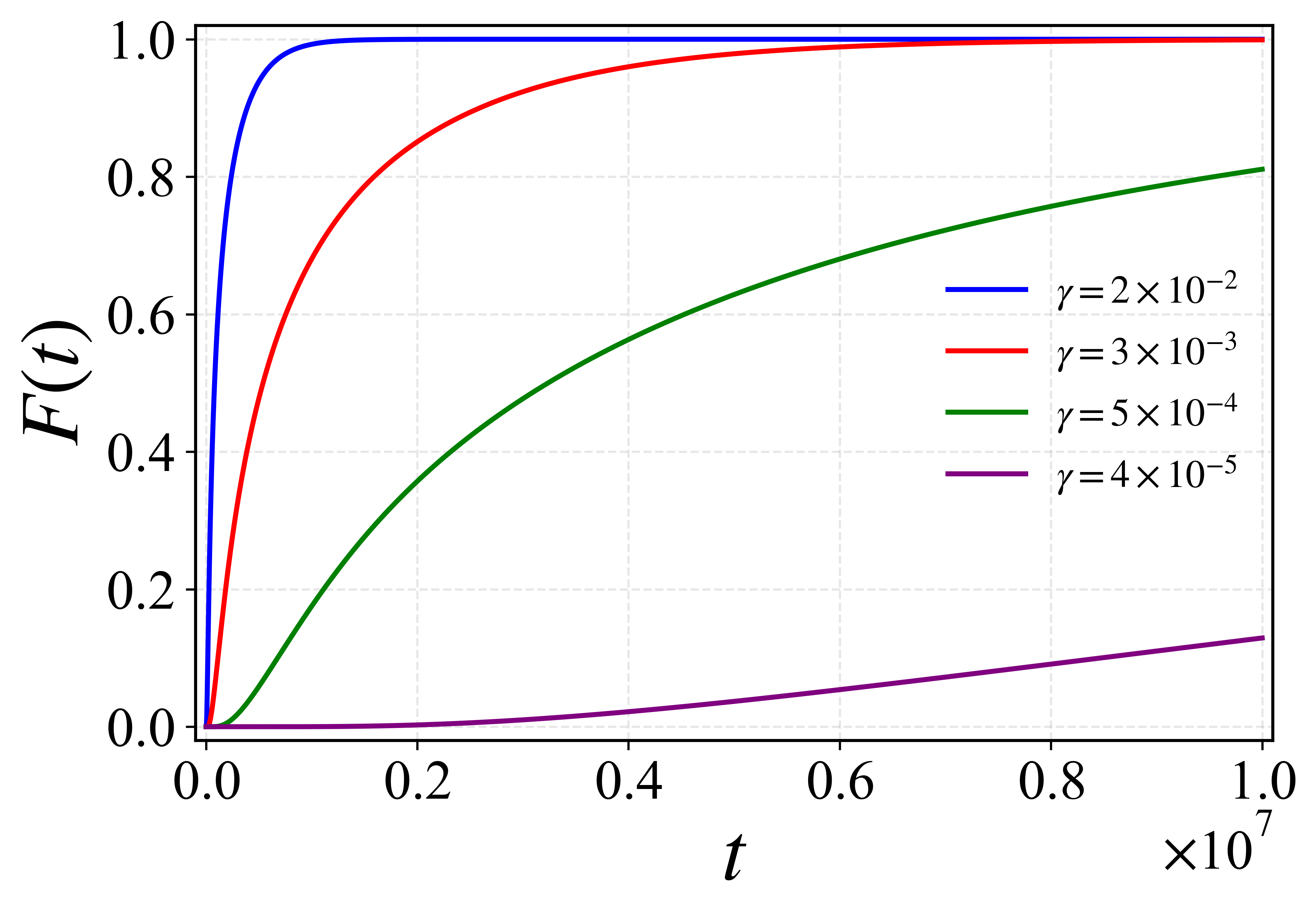}  
    \caption{Fidelity $F(t)$ between the time-evolved correlation matrix ($\mathcal{C}(t)$) and the steady-state matrix for different dissipation strengths $\gamma$.
    }
    \label{fig:Fidelity}
\end{figure}

Figure~\ref{fig:Fidelity} shows the time evolution of the fidelity for different dissipation strengths $\gamma$. For all values of $\gamma$, the fidelity approaches unity at long times, confirming convergence towards the unique steady state. However, the convergence rate depends strongly on the dissipation strength. Larger values of $\gamma$ lead to rapid relaxation and a correspondingly fast approach to the steady state, whereas smaller values of $\gamma$ exhibit significantly slower convergence due to the reduced Liouvillian gap.

These results demonstrate that the appearance of even harmonics is intrinsically connected to the establishment of a current-carrying nonequilibrium steady state. In the weak-dissipation regime, the steady state remains well defined, but increasingly long relaxation times are required before its characteristic symmetry-breaking signatures become visible in the HHG spectrum. The fidelity analysis confirms that the observed crossover is governed by the Liouvillian relaxation dynamics and provides a quantitative measure of the timescales required to reach the steady-state regime.

\section{Summary and outlook} \label{s9}
We have investigated high-harmonic generation in a one-dimensional Su-Schrieffer-Heeger chain coupled to source and sink reservoirs and driven by a  laser field. Using a Lindblad master-equation approach, we demonstrated that the dissipative boundaries generate a nonequilibrium steady state carrying a finite dc current. Although the SSH Hamiltonian remains inversion symmetric, the current-carrying steady-state density matrix does not. As a consequence, the conventional symmetry constraints that suppress even-order harmonics are lifted, allowing even harmonics to emerge without any structural inversion-symmetry breaking of the lattice.

The analysis reveals that the relevant symmetry governing harmonic selection rules in open quantum systems is that of the full Liouvillian rather than the Hamiltonian alone. The boundary gain and loss processes break the inversion symmetry of the steady-state density matrix, leading to a finite current and modifying the nonlinear optical response. We showed that the appearance of even harmonics is directly correlated with the establishment of this nonequilibrium steady state and disappears as the influence of the dissipative driving is reduced.

The steady state was obtained both analytically through a Sylvester-equation formulation of the correlation-matrix dynamics and dynamically through time-dependent Lindblad evolution. The comparison between these approaches allowed us to investigate experimentally relevant  preparation protocols. By analyzing the fidelity between the time-evolved and exact steady-state correlation matrices, we quantified the relaxation dynamics and demonstrated the crucial role of the Liouvillian gap in determining the timescale required to reach the steady state.

Our results identify dissipative transport as a mechanism for generating even harmonics in centrosymmetric systems and establish high-harmonic generation as a sensitive probe of current-carrying nonequilibrium steady states. More broadly, they demonstrate how dissipative dynamics can modify  optical selection rules, opening new opportunities for exploring transport and symmetry breaking in driven open quantum materials.

Several open questions remain. First, the present work focuses on a noninteracting one-dimensional SSH chain, whereas realistic quantum materials often exhibit significant many-body interactions. Understanding how interactions influence current-induced even-harmonic generation represents an important next step. Second, it is essential to determine to what extent the mechanism identified here is specific to one-dimensional systems or instead reflects a more general property of nonequilibrium steady states. In particular, it would be interesting to investigate current-induced symmetry breaking in higher-dimensional materials, where nonequilibrium transport can coexist with nontrivial band topology, Berry-curvature effects, and transverse current responses. Candidate systems include Chern insulators, Haldane models, topological semimetals, and other quantum materials supporting topological edge or surface states.

Third, while the present study employs Markovian Lindblad reservoirs to generate nonequilibrium steady states, a more realistic description of source and sink contacts could be achieved within the nonequilibrium Green's function framework. In particular, Keldysh Green's function techniques naturally incorporate energy-dependent lead self-energies, finite-temperature reservoirs, non-Markovian effects, and self-consistent transport under applied bias. Extending high-harmonic generation calculations to current-carrying nonequilibrium steady states within the Keldysh formalism would establish a direct connection to experimentally relevant quantum devices and nanoscale transport geometries.

More broadly, the present work suggests that harmonic selection rules in open quantum systems are governed by the symmetry of the nonequilibrium density matrix rather than solely by the symmetry of the underlying Hamiltonian. Exploring this principle in interacting, higher-dimensional, and topological systems may reveal new mechanisms for controlling nonlinear optical responses through transport and dissipation. Such developments could establish high-harmonic spectroscopy as a powerful probe of nonequilibrium quantum matter and of the interplay between topology, transport, dissipation, and ultrafast nonlinear optics.

\section*{Acknowledgments}
M.J. is grateful to M. V. Hosseini for useful comments. Funding by the Deutsche Forschungsgemeinschaft (DFG, German Research Foundation) through IRTG 2676/1 ‘Imaging of Quantum Systems’, project no. 437567992 is gratefully acknowledged.


\appendix

\section{Lindblad Equation for Single-Particle Density Matrix}
\label{App_DensityMatrix}

We derive the equation of motion for the single-particle correlation matrix in the presence of arbitrary local particle-gain and particle-loss reservoirs~\cite{nava2023lindblad,wang2024dynamic}. The correlation matrix is defined as
\begin{equation}
\mathcal{C}_{mn}(t)=\mathrm{Tr}\left(c_m^\dagger c_n\rho(t)\right)
=\langle c_m^\dagger c_n\rangle ,
\end{equation}
where $\rho(t)$ is the many-body density matrix and $c_m$, $c_n^\dagger$ are fermionic annihilation and creation operators satisfying the canonical anticommutation relations (CAR),
\begin{equation}\label{CAR}
\{c_i,c_j^\dagger\}=\delta_{ij},\qquad
\{c_i,c_j\}=0,\qquad
\{c_i^\dagger,c_j^\dagger\}=0 .
\end{equation}
The Lindblad equation is
\begin{equation}
\dot{\rho}
=-i[h(t),\rho]
+\sum_k
\left(
L_k\rho L_k^\dagger
-\frac12\{L_k^\dagger L_k,\rho\}
\right).
\end{equation}
We derive the contributions to the correlation matrix from each term separately.

\subsection*{1. Coherent (Hamiltonian) Contribution}
For a quadratic Hamiltonian,
\begin{equation}
h=\sum_{a,b}h_{ab}c_a^\dagger c_b ,
\end{equation}
the coherent contribution is
\begin{equation}
\dot{\mathcal C}_{mn}\big|_{\rm coh}
=
-i\langle[c_m^\dagger c_n,h]\rangle .
\end{equation}
Using the CAR, the commutator of two quadratic fermionic operators is
\begin{equation}
[c_m^\dagger c_n,c_a^\dagger c_b]
=
\delta_{na}c_m^\dagger c_b
-\delta_{bm}c_a^\dagger c_n .
\end{equation}
Therefore,
\begin{align}
[c_m^\dagger c_n,h]
&=
\sum_{a,b}h_{ab}
[c_m^\dagger c_n,c_a^\dagger c_b]
\nonumber\\
&=
\sum_b h_{nb}c_m^\dagger c_b
-\sum_a h_{am}c_a^\dagger c_n .
\end{align}
Taking the expectation value gives
\begin{align}
\dot{\mathcal C}_{mn}\big|_{\rm coh} &= -i\sum_bh_{nb}\mathcal C_{mb} +i\sum_ah_{am}\mathcal C_{an}\nonumber\\
&= i[h^T,\mathcal C]_{mn}.
\end{align}
Hence, in matrix form, and considering that the SSH Hamiltonian has real hopping amplitudes, such that $h=h^\dagger=h^T$, the coherent contribution can be written as
\begin{equation}
\dot{\mathcal C}\big|_{\rm coh}=i[h,\mathcal C].
\label{Ham_form}
\end{equation}

\subsection*{2. Dissipative Contribution}

The contribution from a single Lindblad operator is
\begin{equation}
\mathcal D[\rho]
=
L\rho L^\dagger
-\frac12\{L^\dagger L,\rho\}.
\end{equation}

\subsubsection*{2.1 Particle Loss Reservoirs}

For a loss reservoir attached to site $p$,
\begin{equation}
L_p^{\rm loss}
=
\sqrt{\gamma_l^p}\,c_p ,
\end{equation}
the dissipative contribution becomes
\begin{align}
\dot{\mathcal C}_{mn}^{(p,-)}&=\gamma_l^p
\left(\langle c_p^\dagger c_m^\dagger c_n c_p\rangle-\frac12
\langle\{c_p^\dagger c_p,c_m^\dagger c_n\}\rangle\right).
\end{align}
Using the CAR,
\begin{align}
\{c_p^\dagger c_p,c_m^\dagger c_n\}&=\delta_{pm}c_p^\dagger c_n+\delta_{np}c_m^\dagger c_p\nonumber\\
&\quad+2c_p^\dagger c_m^\dagger c_nc_p ,
\end{align}
and the quartic terms cancel. Thus,
\begin{equation}
\dot{\mathcal C}_{mn}^{(p,-)}=-\frac{\gamma_l^p}{2}\left(\delta_{mp}\mathcal C_{pn}+\delta_{np}\mathcal C_{mp}\right).
\end{equation}

Introducing the local loss matrix
\begin{equation}
\Gamma_p^-=\gamma_l^pP_p,\qquad P_p=|p\rangle\langle p|,
\end{equation}
the total contribution from all loss reservoirs is
\begin{equation}
\dot{\mathcal C}\big|_{\rm loss}=-\frac12\sum_p\left(\Gamma_p^-\mathcal C+\mathcal C\Gamma_p^-\right).
\label{loss_general}
\end{equation}

\subsubsection*{2.2 Particle Gain Reservoirs}

For a gain reservoir coupled to site $q$,
\begin{equation}
L_q^{\rm gain}=\sqrt{\gamma_g^q}\,c_q^\dagger ,
\end{equation}
we obtain
\begin{align}
\dot{\mathcal C}_{mn}^{(q,+)}&=\gamma_g^q\left(\langle c_qc_m^\dagger c_nc_q^\dagger\rangle-\frac12 \langle
\{c_qc_q^\dagger,c_m^\dagger c_n\}\rangle\right).
\end{align}
Using the CAR,
\begin{align}
\{c_qc_q^\dagger,c_m^\dagger c_n\}&=2c_m^\dagger c_n-\delta_{qm}c_q^\dagger c_n-\delta_{nq}c_m^\dagger c_q\nonumber\\
&\quad +2c_q^\dagger c_m^\dagger c_qc_n ,
\end{align}
and again the quartic terms cancel. We obtain
\begin{equation}
\dot{\mathcal C}_{mn}^{(q,+)}= \gamma_g^q
\left(\delta_{mq}\delta_{nq} -\frac12
\delta_{mq}\mathcal C_{qn}-\frac12 \delta_{nq}\mathcal C_{mq} \right).
\end{equation}
Defining
\begin{equation}
\Gamma_q^+ = \gamma_g^qP_q ,
\end{equation}
the contribution from all gain reservoirs is
\begin{equation}
\dot{\mathcal C}\big|_{\rm gain} = \sum_q \left[
\Gamma_q^+ -\frac12 \left( \Gamma_q^+\mathcal C + \mathcal C\Gamma_q^+ \right) \right].
\label{gain_general}
\end{equation}
Combining Eqs.~\eqref{Ham_form}, \eqref{loss_general}, and \eqref{gain_general}, the full equation of motion for the single-particle correlation matrix is
\begin{align}
\dot{\mathcal C}=&i[h,\mathcal C]-\frac12\sum_p \left( \Gamma_p^-\mathcal C + \mathcal C\Gamma_p^- \right) \nonumber\\
&+ \sum_q \left[ \Gamma_q^+ -\frac12 \left( \Gamma_q^+\mathcal C + \mathcal C\Gamma_q^+ \right) \right].
\end{align}
This equation is valid for an arbitrary number of local gain and loss reservoirs. The specific boundary-driven setup used in the main text is obtained by choosing one loss reservoir at $p=1$ and one gain reservoir at $q=2N$ with equal coupling strengths,
$\gamma_l=\gamma_g=\gamma$.

\section{Exact solution via the bi\-orthonormal spectrum of $\mathcal{M}$}\label{AppSS}

We work in the single-particle site basis $\{|j\rangle\}_{j=1}^{2N}$ 
introduced in the main text, with the explicit correspondence 
$\{|j\rangle\} = \{|A_1\rangle,|B_1\rangle,\ldots,|A_N\rangle,|B_N\rangle\}$,
with $\langle i|j\rangle=\delta_{ij}$.
The generator is
\begin{align}
\mathcal{M} &= i h - \tfrac{1}{2}(\Gamma^+ + \Gamma^-), \\ \nonumber 
\Gamma^+ &= \gamma|2N\rangle\langle 2N|,\\ \nonumber 
\Gamma^-&= \gamma|1\rangle\langle 1|.
\end{align}
We assume that $\mathcal{M}$ is diagonalizable and that a complete set of biorthonormal
right and left eigenvectors $\{|u_k\rangle, \langle \tilde u_k|\}$ satisfying
\[
\mathcal{M}|u_k\rangle = \lambda_k |u_k\rangle, \qquad
\langle \tilde u_k| \mathcal{M} = \lambda_k \langle \tilde u_k|,
\]
with bi-orthonormality and completeness
\[
\langle \tilde u_k|u_{k'}\rangle = \delta_{kk'}, \qquad
\sum_k |u_k\rangle\langle \tilde u_k| = \mathbb{I}.
\]
Starting from \(|\psi(0)\rangle\), one can write
\[
|\psi(t)\rangle := e^{\mathcal{M} t}|\psi(0)\rangle
= \sum_k |u_k\rangle\langle\tilde u_k|\psi(0)\rangle e^{\lambda_k t}.
\]
(One obtains this by inserting the resolution of identity \(\sum_k |u_k\rangle\langle\tilde u_k|\) and using \(e^{\mathcal{M} t}|u_k\rangle=e^{\lambda_k t}|u_k\rangle\).)
Using \(\Gamma^+=\gamma|2N\rangle\langle 2N|\)  we get the integrand
\begin{align}
&e^{\mathcal{M} t}\Gamma^+ e^{\mathcal{M}^\dagger t}= \\ \nonumber 
&\gamma\sum_{k,\ell} \langle\tilde u_k|2N\rangle \langle 2N|\tilde u_\ell\rangle^*\, e^{\lambda_k t} e^{\lambda_\ell^* t}\, |u_k\rangle\langle u_\ell|.
\end{align}
Integrate termwise (justified when the integral converges; see below):
\[
\mathcal{C}_{\rm ss}
= \gamma\sum_{k,\ell} \langle\tilde u_k|2N\rangle \langle 2N|\tilde u_\ell\rangle^*
\int_0^\infty e^{(\lambda_k+\lambda_\ell^*)t}\,dt \; |u_k\rangle\langle u_\ell|.
\]
If \(\Re(\lambda_k)<0\) the integral converges and gives
\[
\int_0^\infty e^{(\lambda_k+\lambda_\ell^*)t}\,dt
= -\frac{1}{\lambda_k+\lambda_\ell^*}.
\]
Thus, the exact representation is
\[
\mathcal{C}_{\rm ss} = \gamma \sum_{k,\ell} \frac{\langle\tilde u_k|2N\rangle \langle 2N|\tilde u_\ell\rangle^*}{-(\lambda_k+\lambda_\ell^*)}\; |u_k\rangle\langle u_\ell|.
\]

\section{Correlation-matrix form of the Uhlmann fidelity for pairing-free Gaussian states}\label{App:Fidelity}
This appendix demonstrates that our model's open dynamics naturally keeps the state in the manifold of pairing-free fermionic Gaussian states, and derives the general expression for the Uhlmann fidelity between two such states in terms of their correlation matrices. Specializing this result to the time-evolved state and the steady state reduces it to Eq.~\eqref{eq:fidelity} in the main text.

Throughout this, we consider fermionic modes satisfying the canonical anticommutation relations of Eq.~\eqref{CAR}, and denote the normal and anomalous correlation matrices as
\begin{align*}
&C_{mn}=\langle c_m^{\dagger}c_n\rangle,
\quad C=C^{\dagger},\quad 0\le C\le\mathbf{1}, \\
&\quad A_{mn}=\langle c_mc_n\rangle, \quad A^\top=-A.
\end{align*}
 
We recall the standard facts used below. A state is Gaussian precisely when its correlators
factorize by Wick's theorem, in which case it is fixed entirely by the pair $(C,A)$ \cite{Bravyi_2005_Lagrangian,araki_quasifree_1970}. With a Lindblad equation with quadratic Hamiltonian and linear jump operators in the ladder
operators, the dynamics preserves Gaussianity  \cite{brodier_symplectic_2004,ferraro_gaussian_2005,graefe_lindblad_2018-1,christie_quantum-jump_2022}
\begin{equation}
    \rho(t=0)\in\mathcal G\ \Rightarrow\ \rho(t)=e^{\mathcal{L}t}\rho(t=0)\in\mathcal G,\quad \forall t\ge0,
\end{equation}
and expectation values evolve through the adjoint generator.
Here $\mathcal{G}$ is the set of fermionic Gaussian states and $\mathcal{L}$ is the Lindblad generator.

We compute the equation of motion of the anomalous block by inserting $\mathcal O=c_mc_n$ into
\begin{equation}\label{eq:adjoint}
\mathcal{L}_t^{\dagger}[\mathcal{O}]=i\,[H,\mathcal{O}]+\sum_k\Big(L_k^{\dagger}\,\mathcal{O}\,L_k-\tfrac12\big\{L_k^{\dagger}L_k,\mathcal{O}\big\}\Big)
\end{equation}
With these fixed conventions fixed, we will derive the equations of motion for the anomalous part used throughout what follows.
 
\subsection{Vanishing contribution of the pairing sector}
 
\paragraph{Coherent contribution.}

The averaged contribution of the coherent part is
\begin{equation}
\begin{aligned}
i\langle[H,c_mc_n]\rangle &=-i\!\sum_b\big(h_{mb}A_{bn}+h_{nb}A_{mb}\big)\\
&=-i\big(hA+Ah^{\mathsf T}\big)_{mn} .
\end{aligned}
\end{equation}

\paragraph{Loss contribution.}
For $L=\sqrt{\gamma_{\ell p}}\,c_p$, the operator $c_mc_n$ contains no creation operator on the loss site and hence, the first term of the dissipation part
reduces to $n_p\,c_mc_n$ where $n_p$ is the number operator.
Normal-ordering the anticommutator part then results in a contribution
\begin{equation}
\dot A_{mn}\big|_{\ell}
=-\frac{\gamma_{\ell p}}{2}\big(\delta_{pm}+\delta_{pn}\big)A_{mn}.
\end{equation}
 
\paragraph{Gain contribution.}
For $L=\sqrt{\gamma_{gq}}\,c_q^{\dagger}$, normal ordering the first term of the
dissipator against the anticommutator part yields
\begin{equation}
\dot A_{mn}\big|_{g}
=-\frac{\gamma_{gq}}{2}\big(\delta_{qm}+\delta_{qn}\big)A_{mn},
\end{equation}
 Summing the three contributions gives the closed equation of motion with the form 
 \begin{equation}
     \dot A=-i(hA+Ah^{\mathsf T})-\tfrac12(\Gamma A+A\Gamma^{\mathsf T}) ,\qquad\Gamma=\Gamma_\ell+\Gamma_g
 \end{equation}
 which  every term have exactly one factor of $A$, and unlike $\dot C$, no state-independent source survives.
 
Therefore it has schematically the form $\dot A=\mathcal K(t)[A]$ for a linear map $\mathcal K(t)$. An initial condition $A(0)=0$ propagates 
\begin{equation}
    A(t)=0  , \qquad \forall t\ge0
\end{equation}
that fits our model in which we used the half-filling initial state. 

\subsection{From correlation matrices to fidelity}

For pairing-free Gaussian states, the Uhlmann fidelity can be expressed entirely in terms of the single-particle correlation matrices. To derive this result, we first recall the modular Hamiltonian  (also known as the entanglement Hamiltonian or fictitious Hamiltonian) representation of a Gaussian state \cite{Peschel_2003_Calculation,Peschel_2009_Reduced}.
For a pairing-free Gaussian state satisfying $0<C<\mathbf1$,
\begin{equation}\label{eq:modular}
\rho_i
=
\frac{1}{Z_i}
e^{-c^\dagger\mathsf h_i c},
\
C_i
=
\left(e^{\mathsf h_i}+\mathbf1\right)^{-1},
\
Z_i
=
\det(\mathbf1-C_i)^{-1}.
\end{equation}
Since
\[
\mathsf h_i
=
\ln\!\left[(\mathbf1-C_i)C_i^{-1}\right]
\]
diverges when an occupation eigenvalue approaches $0$ or $1$, it is convenient to introduce the transfer matrix
\begin{equation}
X_i
:=
e^{-\mathsf h_i}
=
C_i(\mathbf1-C_i)^{-1}.
\end{equation}
We adopt the squared convention for the Uhlmann fidelity~\cite{Uhlmann_Fid},
\begin{equation}
F(\rho_1,\rho_2)
=
\left(
\operatorname{Tr}
\sqrt{\sqrt{\rho_1}\,\rho_2\,\sqrt{\rho_1}}
\right)^2.
\end{equation}
The square root of a Gaussian state is again Gaussian with the halved modular Hamiltonian 
\[
\sqrt{\rho_1}=Z_1^{-1/2}e^{-c^{\dagger}\frac{h_1}{2}c},
\]
and products of Gaussian exponentials can be expressed in closed form. The correspondence $e^{-c^{\dagger}Kc}\mapsto e^{-K}$ maps a Gaussian
operator to its associated single-particle transfer matrix and forms a group homomorphism. Consequently,
\begin{equation}
G:=\sqrt{\rho_1}\,\rho_2\sqrt{\rho_1}=(Z_1Z_2)^{-1}e^{-c^{\dagger}\gamma c},
\end{equation}
where
\begin{equation}
e^{-\gamma}=e^{-h_1/2}e^{-h_2}e^{-h_1/2}=X_1^{1/2}X_2X_1^{1/2}=:M .
\end{equation}
 Collecting all factors and applying the trace yields 
\begin{equation}
\operatorname{Tr}\sqrt{G}=(Z_1Z_2)^{-1/2}\det\!\big(\mathbf{1}+M^{1/2}\big).
\end{equation}
Finally, for pairing-free Gaussian states with $0<C_i<\mathbf{1}$,
\begin{equation}\label{c-fidelity}
\begin{aligned}
    F(\rho_1,\rho_2)=\det\!\big[(\mathbf{1}-C_1)(\mathbf{1}-C_2)\big]\\
\times\det\!\Big[\mathbf{1}+\big(\sqrt{X_1}X_2\sqrt{X_1}\big)^{1/2}\Big]^{2}.
\end{aligned}
\end{equation}
Equation~\eqref{c-fidelity} provides a general expression for the Uhlmann fidelity between any two pairing-free Gaussian states. In the present work, we use it to quantify the overlap between the instantaneous state $\rho(t)$ and the nonequilibrium steady state $\rho_{\mathrm{ss}}$. Identifying
\[
C_1=\mathcal C(t),
\qquad
C_2=\mathcal C_{\mathrm{ss}},
\]
the fidelity becomes
\begin{equation}
\begin{aligned}
F(t)
\equiv
&F\left(\rho(t),\rho_{\mathrm{ss}}\right)=
\det\!\Big[(\mathbf1-\mathcal C(t))
(\mathbf1-\mathcal C_{\mathrm{ss}})\Big]\\
&\times \det\left[
\mathbf1+
\left(
\sqrt{X(t)}
X_{\mathrm{ss}}\,
\sqrt{X(t)}
\right)^{1/2}
\right]^2,
\end{aligned}
\end{equation}
where
\begin{equation}
X(t)
=
\mathcal C(t)
\bigl(\mathbf1-\mathcal C(t)\bigr)^{-1},
\qquad
X_{\mathrm{ss}}
=
\mathcal C_{\mathrm{ss}}
\bigl(\mathbf1-\mathcal C_{\mathrm{ss}}\bigr)^{-1}.
\end{equation}
Thus, the fidelity is evaluated entirely from the time-dependent and steady-state correlation matrices, without explicitly constructing the corresponding many-body density matrices.


\bibliography{References} 


\end{document}